\documentclass[12pt,preprint]{aastex}

\shorttitle{Chandra observations of X-ray weak Quasars}
\shortauthors{Risaliti et al.}

\begin{document}

\title{A Chandra mini-survey of X-ray weak quasars 
    }

\author{G. Risaliti\altaffilmark{1,2}, M. Elvis\altaffilmark{1} 
    }

\author{
R. Gilli\altaffilmark{2}, M. Salvati\altaffilmark{2}}
\email{grisaliti@cfa.harvard.edu}

\altaffiltext{1}{Harvard-Smithsonian Center for Astrophysics, 60 Garden st. 
Cambridge, MA 02138 }
\altaffiltext{2}{INAF - Osservatorio Astrofisico di Arcetri, L.go E. Fermi 5,
I-50125 Firenze, Italy}

\begin{abstract}
We present {\em Chandra} observations of 18 spectroscopically selected quasars,
already known to be X-ray weak from previous ROSAT observations. 
All the sources but one are detected by {\em Chandra}, and spectral analysis
suggests that most of them are intrinsically underluminous in the X-rays (by a
factor from 3 to $>100$). 
These objects could represent a large population of quasars with
a Spectral Energy Distribution different from that of standard blue
quasars. We discuss the possibility that a significant fraction of the obscured AGN needed
in Synthesis models of the X-ray background could be instead optically
broad-line, X-ray weak quasars.
\end{abstract}


\section{Introduction}

The blue optical spectrum
has normally been used to distinguish quasars 
from stars and normal galaxies
in optical
surveys (BQS, Schmidt \& Green 1986; LBQS, Foltz et al.
1990; 2dF, Boyle et al. 2000). As a
consequence, our knowledge of quasars is by definition limited to ``blue''
quasars. 
However, many different quasar SED could exist, 
undiscovered because of the limits of the available
instruments and selection criteria (Elvis 1992).
The X-ray properties of optically selected quasars show
a similar homogeneity: the 1-10 keV spectrum is well represented by a
power law with photon index $\Gamma\sim1.8-2$ (Laor et al. 1997, Reeves \& Turner
2000); the average optical to X-ray slope  for optically 
selected samples is $\alpha_{\rm OX}=1.55$ (Laor et al. 1997\footnote{ 
$\alpha_{\rm OX}$ is defined as the spectral index of a power law connecting the 
points at 2500~\AA~and at 2 keV (rest frame) of the quasar SED in the ($\nu,f_\nu$) plane.})
Only a minority of objects ($<10\%$) are significantly weaker
($\alpha_{\rm OX}>1.8$) than the
average in the 0.5-2~keV X-ray band (Yuan et al. 1998). 

When quasars are searched with selection criteria other than the U-B
color, different properties emerge. Webster et al. (1995) found a large
population of red broad-line quasars in radio surveys. Kim \& Elvis (1998)
discovered red quasars in soft X-ray selected samples. Most strikingly,
the 2MASS survey (Skrutskie et al. 1997) 
found a large number of red quasars, whose density
is of the same order of that of local color-selected quasars (Cutri et al.
2001). A {\em Chandra} survey of these objects revealed X-ray properties completely
different from ``normal'' quasars: the X-ray emission is much weaker and the
spectra are flatter, suggesting that these objects suffer significant
absorption by circumnuclear gas (Wilkes et al. 2002). The existence of these
populations is important. Synthesis models for the X-ray (Comastri et al.
1995, Gilli, Salvati \& Hasinger 2001) and FIR (Risaliti, Elvis \& Gilli 2002)
backgrounds depend sensitively on knowing the true AGN population. Similarly,
the total accretion luminosity of the Universe (Fabian \& Iwasawa 1999) and
the average efficiency of black hole accretion (and hence spin, Elvis,
Risaliti \& Zamorani 2002) depend primarily on these ``hidden'' populations.

Another indication of unexpected X-ray properties of non-color selected
quasars comes from the work of Risaliti et al. 2001 (hereafter R01), where a sample of 
spectroscopically selected quasars of the Hamburg survey (Hagen et al. 1995)
has been cross-correlated with the WGA Catalogue of ROSAT pointed observations
(White, Giommi \& Angelini 1995). More than half of the resulting sample
is underluminous in the X-rays, by a factor from $\sim 5$ to $>100$.
Interestingly, most of these objects are somewhat redder 
than ``normal'' quasars ($\Delta(B-R)\sim~1$ vs. $\Delta(B-R)\sim~0.5$, R01), and
would have been probably missed in standard color-based surveys, since we
expect their U-B color also to be redder than the average of standard blue
quasars.
Almost all the objects in this sample were not detected by ROSAT, therefore
the claim on their X-ray weakness is based on upper limits. As a consequence,
nothing is known about their X-ray spectral properties.

The subarcsecond beam size of the {\em Chandra} mirrors 
(van Speybroeck et al.
1997) and the large collecting area
endows {\em Chandra} with far greater sensitivity than ROSAT. Hence, a
{\em Chandra} survey can explore the X-ray sample of quasars
far better than ROSAT.
Here we present the results of {\em Chandra} observations of 18 objects selected
from the X-ray weak sample of R01.
\section{Sample selection and observations}
The parent sample was obtained by R01
from the cross-correlation of the Hamburg Quasar Survey (hereafter HS, 
Hagen et al. 1995) with the WGA
Catalogue. The HS sample consists of 397 quasars, with redshift between 0 and
3, and limiting magnitude B$\sim 19-19.5$. The selection criteria are either
the standard U-B colour or the presence of broad emission lines (or both) in grism
optical spectra. In this way it is possible to discover objects with 
intrinsically red continua, or with moderate extinction.

The R01 sample contains 85 sources, of which only 31 were
detected by ROSAT. In Fig.~1 we show the distribution of X-ray to
optical ratios for this sample, compared with the one of PG quasars.
We adopted an index, defined in R01 as
I$_{\rm OX}=\frac{20-B}{2.5}-\log~\phi$, where $\phi$ is the ROSAT 0.5-2.4 keV count rate.
The reason for using this new index, instead of $\alpha_{OX}$, is that 
it uses the observed optical data since, with these redder quasars,
an extrapolation to 
2500~\AA~may be problematic.
For quasars with a standard SED, I$_{\rm OX}\sim
2.6\alpha_{\rm OX}-1.3$. \\ 
The detected sources in R01 have a rather normal X-ray to optical
ratio, while the 54 non-detections are X-ray weaker than most PG quasars. 
The dashed vertical line in Fig. 1 (I$_{\rm OX}=3.2$) represents the value at which
the underluminosity in the X-ray is a factor of 5 with respect to the average
of PG quasars. In the following we refer 
to I$_{\rm OX}>3.2$ as ``X-ray weak'' sources.

We randomly selected 17 sources from the X-ray weak half of the sample.
As a control sample we included three HS quasars not detected by ROSAT
but with upper limits on I$_{\rm OX}$ lower than 3.2. 

As shown by R01, there is a strong correlation between the optical O-E
color, obtained from the POSS plates, and the X-ray to optical ratio. As a 
consequence, our 17 X-ray weak sources, being a representative sub-sample
of the X-ray weak quasars of R01, are automatically also a representative sub-sample of the
red quasars present in the Hamburg Survey, i.e. those objects that would have not been
selected by optical color-based surveys.

18 out of the 20 sources of the sample have been observed with the ACIS-S
detector on {\em Chandra} (Weisskopf et al. 2001) in the
year 2002. 
Out of these 18 objects, 16 are from the X-ray weak group, and the remaining 2
are from the small group of 3 control sources.
The observing times vary from 4 to 10 ksec, given the B magnitude of
the sources. The observing times were chosen in order to have the
same lower limit on $\alpha_{\rm OX}$ for all the sources, in case of non-detection
with {\em Chandra}. 
All the sources but one have been detected. The net source counts range from a few
tens counts to $\sim 1000$ for a few bright objects (Tab. 1).

\section{Analysis and Results}

The data were analyzed using the latest ACIS calibrations provided by the 
{\em Chandra} X-ray Center. 
A correction was applied to the response matrices to account for 
the low energy quantum efficiency
degradation of the ACIS detector\footnote{URL:http:$//$asc.harvard.edu$/$ciao$/$threads$/$apply\_acisabs$/$index.html}. 
We analyzed the ACIS spectra using a simple model, consisting of an 
absorbed power law. 
Both the photon index, $\Gamma$ and the absorbing column
density $N_H$ were left free. (Table 1)
In Fig. 2 we show the results for the 16 objects of the X-ray weak sample, compared with the
spectra estimated from the B magnitudes, assuming the normal UV-selected value of 
$\alpha_{\rm OX}=1.55$ (shaded
band in each panel of Fig. 2).
We calculated $\alpha_{\rm OX}$ deriving the monochromatic flux at 2500~\AA~from the B
magnitude, assuming a spectral shape $f_\nu \propto \nu^{-0.5}$. 
This is a conservative assumption because the optical colors 
of these objects are on average redder than in normal quasars. Since for all but three
sources the B magnitude central wavelength ($\lambda_B=$4400~\AA) is greater than the
redshifted 2500 \AA~wavelength, using flatter optical spectra would imply higher
extrapolations of $f_\nu$ at 2500 \AA, and so even higher values of $\alpha_{\rm OX}$. 
The rest-frame 2 keV flux
was directly measured from the spectrum. 
The results can be summarized as follows:\\
$\bullet$ 12 out of the 16 X-ray weak sources are confirmed to be
extremely faint in the X-rays, with $\alpha_{\rm OX}$ ranging from 1.7 to 2.3.
Another object, HS 1417+4522, is only marginally weaker than the average at 2 keV 
($\alpha_{\rm OX}=1.58$), but is significantly weaker than normal AGNs at higher energies
(Fig. 2).
The spectra are on average slightly flatter than the canonical $\Gamma=1.8$ (Risaliti 2002):
we computed a stacked spectrum of these 13 sources and we obtained 
an average photon index $\Gamma=1.5$.\\
$\bullet$ 3 out of 16 objects have a ``normal'' $\alpha_{\rm OX}$, significantly
higher than the estimate from the ROSAT upper limit. This implies strong
variability (of at least a factor $\sim 10$ in the 0.5-2 keV band).
Interestingly, these three objects are the only ones 
in our sample at redshift lower than unity. We will further discuss these sources
in a forthcoming paper.\\
$\bullet$ The two ``control sources'' both show a ``normal'' X-ray spectrum.
As a consequence, there is no indication that a significant fraction of the
sources in the left part of the histogram in Fig. 1 are X-ray weak. Our best
estimate of the fraction of X-ray weak quasars in the Hamburg quasars remains
the one inferred from the ROSAT observations, i.e. $\sim 50\%$.
\section{Discussion}
Since 13 out of 16 objects are confirmed to be X-ray weak by {\em
Chandra}
observations, 
the fraction of X-ray weak sources in the
parent sample of R01 is as high as 13/16 of 50\%, i.e. $\sim~40$\%.
Since the space density of these sources is of the
same order of that found in color-selected surveys with similar limiting
magnitudes, these objects represent a significant part of the AGN population.
From the X-ray point of view, they appear to be completely different from 
standard blue quasars both in $\alpha_{\rm OX}$ and $\Gamma$. 
In principle, two interpretations are possible for the above results:
the sources can be either (1) heavily absorbed or (2) intrinsically X-ray weak.

(1) If absorption plays a crucial role, the observed radiation could be due to
warm scattering and/or cold reflection\footnote{Note that diffuse emission from the
host galaxies is expected to give little contribution, the observed luminosity
being higher than 10$^{44}$ erg s$^{-1}$.} while the intrinsic emission would be
absorbed by a column density $N_H>10^{24}$~cm$^{-2}$.
These objects are somewhat redder than normal ``blue
quasars'' 
and are all broad line quasars. We would then have an unlikely
column density distribution, with all the optically selected blue quasars
having
$N_H<10^{21}$~cm$^{-2}$, all the redder quasars having
$N_H>10^{24}$~cm$^{-2}$, and nothing in between.
This is an argument favoring the alternative hypothesis 
of intrinsic X-ray weakness.

(2) If the quasars are intrinsically X-ray weak, the accretion disk/corona
system could be in a different state than in normal quasars. For example, 
a weaker corona would naturally produce a weaker X-ray emission. 
A widely accepted model is that an X-ray emitting corona is generated by the 
Magneto-Rotational Instability (MRI, Balbus \& Hawley 1991).
This MRI generates the viscosity in the accretion disk, so if MRI is 
ineffective little energy should be liberated and the UV continuum and 
the emission lines should be weak. Yet most
most of our objects
have been selected through the CIV 1549~\AA~line, and therefore it is unlikely
that they have weak ionizing continua. This paradox clearly needs investigating 
theoretically. Perhaps MRI does not produce disk viscosity or the bulk of the X-rays 
in normal quasars have another origin.\\
Our results show that the current view of the X-ray
properties of quasars could be strongly biased  by the optical selection 
towards X-ray loud and
steep-spectra objects. This is supported by the fact
that the other known red quasars have properties similar to our objects (see
for example Wilkes et al. 2001 for {\em Chandra} observations of 2MASS quasars).

As can be seen from Table 1, the X-ray luminosities of our sources are in the range 
10$^{44}-10^{45}$ erg~s$^{-1}$, only slightly higher
than the typical luminosities where the bulk of the X-ray background is
made, according to synthesis models (Gilli et al. 2001). Also, the 
average spectral 
properties ($\Gamma_{AV}=1.5$) are close to those needed by these models. 
It could well be that a fraction of the sources predicted to have N$_H\sim
10^{22}-10^{23}$~cm$^{-2}$, used by current synthesis models are instead 
intrinsically weak, flat spectrum sources, with normal broad quasar emission lines.
If this is the case, the optical/infrared counterparts of these objects would be 
completely different from those of the standard type 2 AGNs: the optical emission 
would not be obscured, but could be redder than standard blue quasars, and therefore less
readily distinguished from stellar emission. 
Another important difference with respect to the standard view is that
the direct optical/UV emission would not be reprocessed 
at mid/far IR wavelengths.
This would significantly lower the expected contribution of AGNs to the far IR background (Risaliti,
et al. 2002) and allowing a larger population of such quasars to be present.

A key test for this hypothesis will be the comparison between the X-ray and optical properties
of the sources with X-ray flux $\sim 10^{-15}$ erg~s$^{-1}$~cm$^{-2}$ in the 
{\em Chandra} Deep Surveys. These objects have luminosities of the order of 10$^{43}-10^{44}$~erg~s$^{-1}$, 
where the bulk of the XRB is made, and their observations have enough S/N to distinguish
between intrinsically weak and absorbed spectra. This will make clear whether 
the X-ray sources described in this work are also common at lower luminosities. 

Another important step to improve the understanding of this class of sources, 
will be the study of their optical and near-IR spectra:
since they are completely different from 
normal blue quasars in the X-rays, they could also have quite different
optical and near-IR SEDs. To explore this issue, we are undertaking 
optical and near-IR observations of several of the sources in our sample at
4-meter class telescopes. 
\acknowledgments

We are grateful to Alexey Vikhlinin for useful comments, and to the referee,
Dr. R. Cutri, for a careful reading of the manuscript.
This work was partially supported by NASA Grant GO2-3142X.

\begin{deluxetable}{lccccccc}
\tablecaption{Data Analysis Results}
\tablehead{
\colhead{Source}           & \colhead{z}      &
\colhead{Exp. time\tablenotemark{a}} & \colhead{Counts} &
\colhead{$\alpha_{\rm OX}$\tablenotemark{b}  }  &
\colhead{$\Gamma$\tablenotemark{c}} &
\colhead{N$_H$\tablenotemark{d}} &
\colhead{L(2-10 keV)\tablenotemark{e}} }
\startdata
HS 0017+2116&2.02 &10130& 62  & 1.77 &  1.79$^{+0.52}_{-0.36}$ & $<0.81$     & 43  \\
HS 0810+5157&0.38 &6940 & 988 & 1.40 &  1.56$^{+0.15}_{-0.07}$ & $<0.14$     & 2.7 \\
HS 0830+1833&2.27 &6480 & 63  & 1.83 &  1.91$^{+0.53}_{-0.33}$ & $<0.65$     & 4.9 \\
HS 0848+1119&2.62 &6120 & 47  & 1.76 &  1.39$^{+0.53}_{-0.38}$ & $<3.20$     & 10.7\\
HS 0854+0915&1.05 &3760 & 37  & 2.07 &  0.67$^{+0.60}_{-0.53}$ & ---         & 1.5 \\
HS 1036+4008&1.96 &6060 & 51  & 2.12 &  1.04$^{+0.66}_{-0.56}$ & ---         & 2.5 \\
HS 1111+4033&2.18 &9760 & 168 & 1.78 &  2.07$^{+0.28}_{-0.25}$ & $<0.43$     & 11.3\\
HS 1202+3538&2.28 &6760 & 52  & 1.79 &  2.28$^{+1.18}_{-0.50}$ & $<2.32$     & 3.2 \\
HS 1229+4807&1.37 &6750 & 100 & 1.58 &  2.17$^{+0.38}_{-0.31}$ & $<0.15$     & 2.9 \\
HS 1237+4756&1.55 &4750 & 393 & 1.38 &  1.55$^{+0.22}_{-0.09}$ & $<0.34$     & 12.1\\
HS 1415+2701&2.50 &8720 &$<$15& $>$2.3 &  ---                  & ---         & $<$3\\
HS 1417+4722&2.21 &7760 & 132 & 1.58 &  2.19$^{+0.46}_{-0.28}$ & $<1.55$     & 12.0\\
HS 1422+4224&2.21 &5960 & 157 & 1.71 &  2.45$^{+0.33}_{-0.44}$ & $<1.10$     & 9.4 \\
HS 1824+6507&0.30 &6950 & 733 & 1.47 &  1.80$^{+0.14}_{-0.12}$ & $<0.11$     & 1.5 \\
HS 1939+7000&0.12 &4970 &1103 & 1.54 &  0.95$^{+0.16}_{-0.07}$ & $<0.02$     & 0.74\\
HS 2135+1326&2.29 &5990 &  53 & 1.88 &  1.98$^{+0.83}_{-0.63}$ & $<2.25$     & 5.8 \\
HS 2146+0428&1.32 &7610 & 170 & 1.73 &  1.88$^{+0.39}_{-0.30}$ & $<0.55$     & 4.1 \\
HS 2251+2941&1.57 &7110 & 35  & 1.97 &  1.50$^{+0.56}_{-0.52}$ & $<2.57$     & 1.5 \\
\enddata
\tablenotetext{a}{Exposure time in seconds.}
\tablenotetext{b}{$\alpha_{\rm OX}$ obtained using the best fit model.}
\tablenotetext{c}{Photon index.}
\tablenotetext{d}{Column density in units of 10$^{21}$~cm$^{-2}$.}
\tablenotetext{e}{2-10 keV luminosity in units of 10$^{44}$ erg s$^{-1}$.}
\end{deluxetable}

\begin{figure}
\plotone{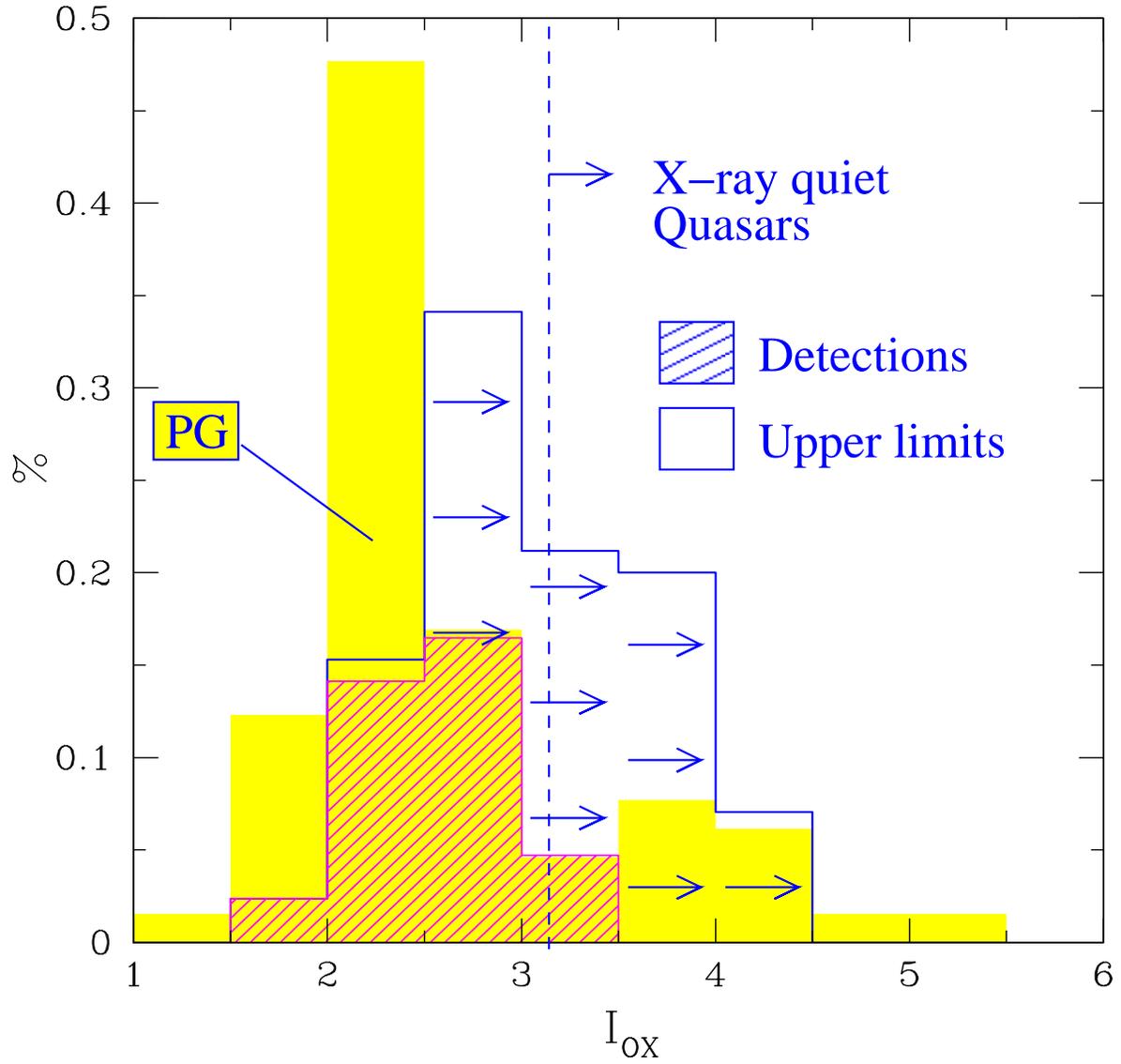}
\figcaption{Optical to X-ray ratio as inferred from ROSAT observations
for the color-selected PG quasars (shaded hystogram) and the sample of R01.
$I_{\rm OX}$ is a logarithmic measure of the ratio between optical (B band)
and soft X-ray flux.}
\end{figure}

\begin{figure}
\begin{center}
\plotone{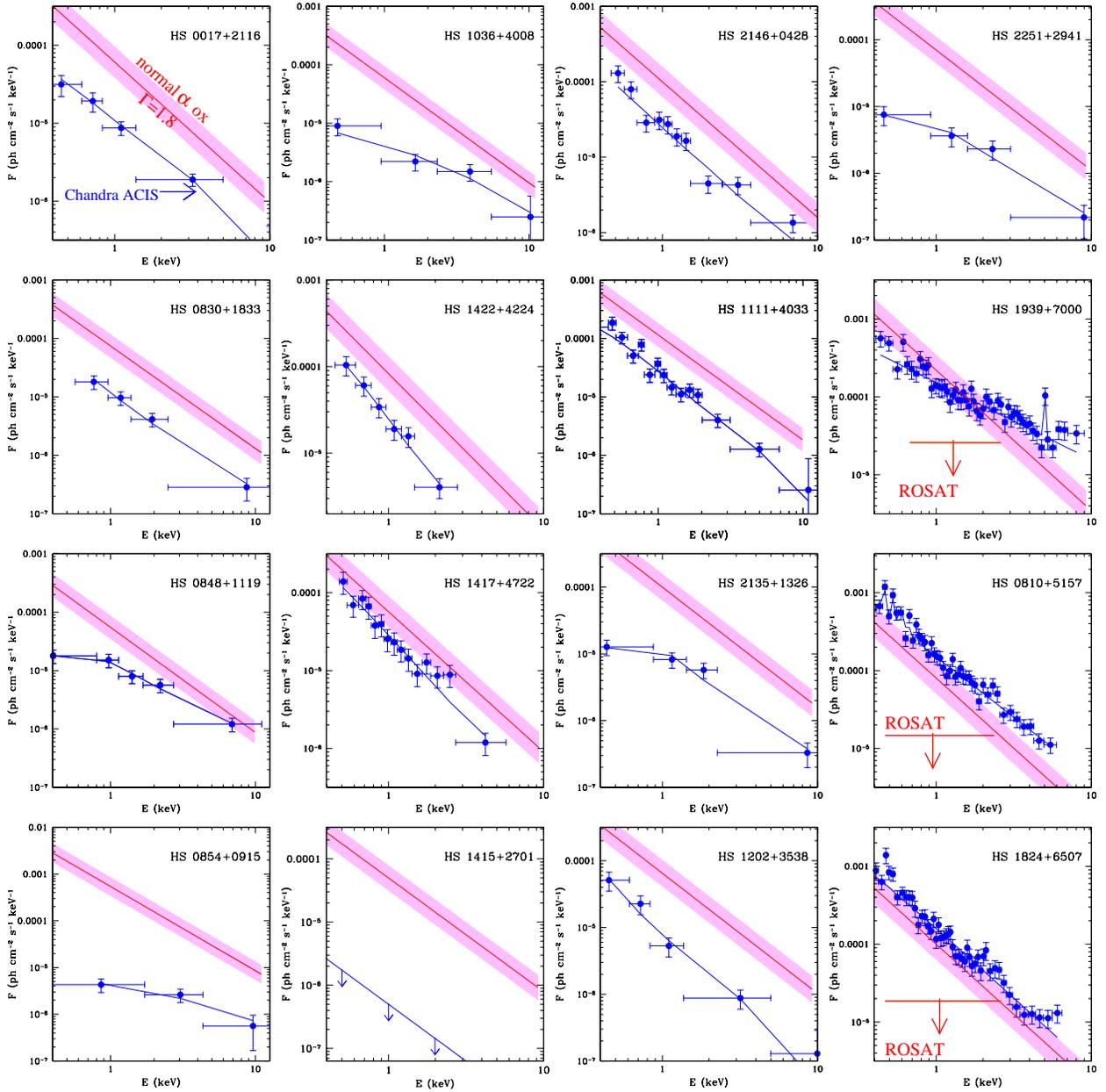}
\figcaption{Chandra spectra of the sample of X-ray weak quasars. 
The shaded region represent the X-ray spectrum expected
assuming $\alpha_{\rm OX}=1.55$ and $\Gamma=1.8$. ROSAT 
upper limits are shown for the three sources having fluxes significantly
higher than in ROSAT observations.}
\end{center}
\end{figure}

\begin{figure}
\plotone{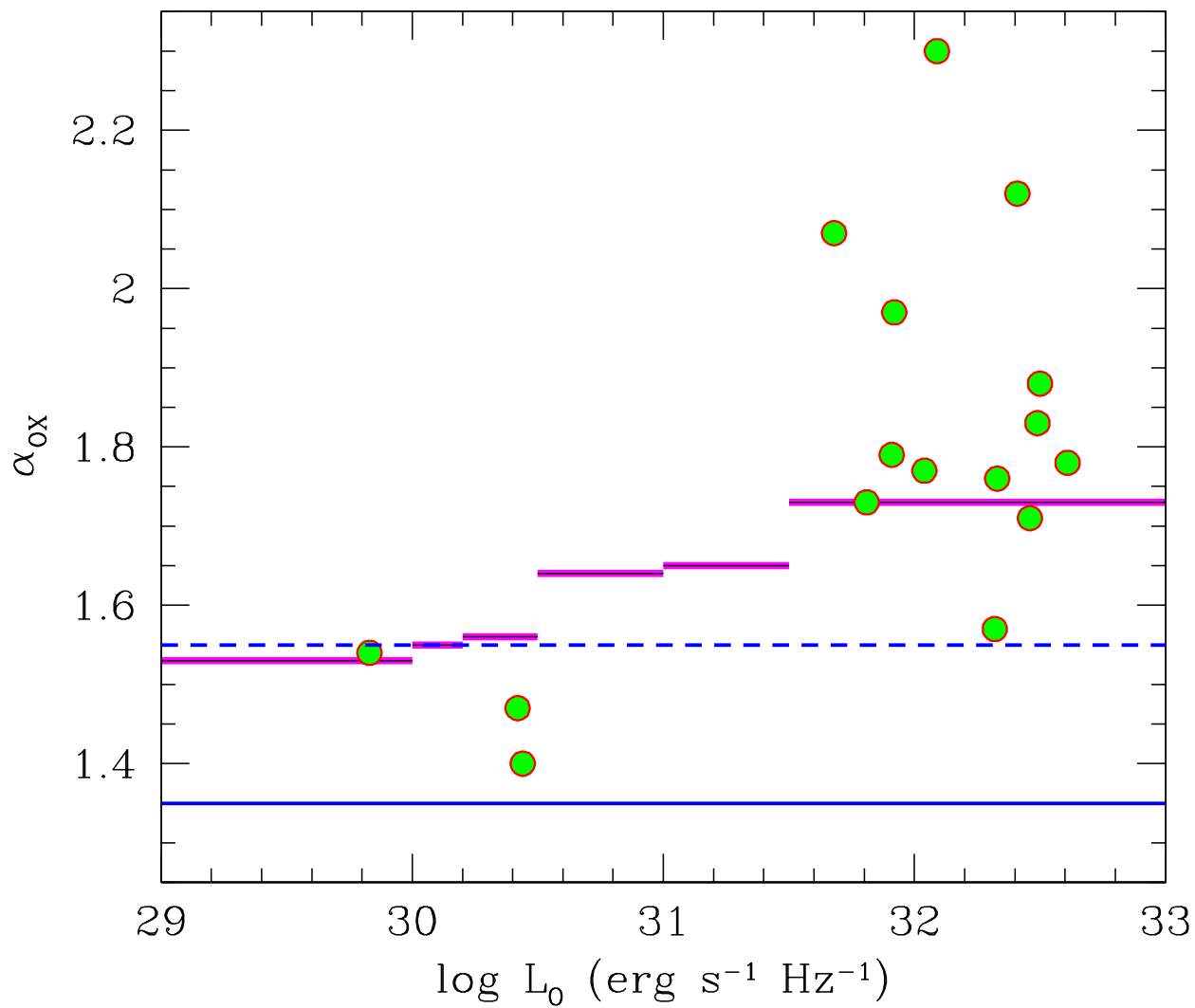}
\figcaption{$\alpha_{\rm OX}$ versus luminosity compared with (a) the average values
found by Yuan et al. (1998) for a sample of $\sim~1000$ optically selected
quasars (shaded lines), (b) the average $\alpha_{\rm OX}$ of PG quasars (dashed line) and (c)
the average $\alpha_{\rm OX}$ of X-ray selected quasars, according to Elvis et al. 1994 (bottom 
continous line). 
The three low-luminosity objects are those with {\em Chandra} fluxes
significantly higher than ROSAT upper limits.
}
\end{figure}

\end{document}